\documentclass{article}
\usepackage{spconf,amsmath,graphicx}

\usepackage[dvipsnames]{xcolor}
\usepackage{hyperref}
\usepackage{graphicx}
\usepackage{comment}

\title{Speaker-Independent Acoustic-to-Articulatory Speech Inversion}
\address{
  $^1$University of California, Berkeley, $^2$Carnegie Mellon University, $^3$University of Southern California}
\begin{document}
%
\maketitle
\begin{abstract}

To build speech processing methods that can handle speech as naturally as humans, researchers have explored multiple ways of building an invertible mapping from speech to an interpretable space. The articulatory space is a promising inversion target, since this space captures the mechanics of speech production. To this end, we build an acoustic-to-articulatory inversion (AAI) model that leverages self-supervision to generalize to unseen speakers. Our approach obtains 0.784 correlation on an electromagnetic articulography (EMA) dataset, improving the state-of-the-art by 12.5\%. Additionally, we show the interpretability of these representations through directly comparing the behavior of estimated representations with speech production behavior. Finally, we propose a resynthesis-based AAI evaluation metric that does not rely on articulatory labels, demonstrating its efficacy with an 18-speaker dataset.

\end{abstract}
\section{Introduction}
\label{sec:intro}

Interpretable representations of speech waveforms are valuable for tasks like diagnosing voice disorders~\cite{lu2022characteristics} and building generalizable, controllable speech synthesizers~\cite{choi2021neural, polyak21facebookresynthesis, anumanchipalli2019speech}. Articulatory speech processing is a promising direction for making speech representations interpretable, since methods in this direction aim to directly model information about vocal tract physiology~\cite{fant1991articulatorysynthesis, rubin1981articulatorysynthesis, scully1990articulatorysynthesis}. Specifically, articulatory representations directly describe the movement of articulators, e.g., the jaw, lips, and tongue. 

Currently, building speaker-independent speech processing models with articulatory data remains challenging, since such data is limited. Most articulatory datasets~\cite{wrench1999mocha, wrench2000multi, richmond2011mngu0, Tiede2017QuantifyingKA} contain only a few hours of high-quality speech, much less than popular non-articulatory ones~\cite{zen2019libritts}. This is mainly since the equipment and processes used to acquire articulatory labels are expensive~\cite{seneviratne2019multi}. A promising, cost-effective alternative to manually collecting articulatory data is acoustic-to-articulatory inversion (AAI), which aims to automatically estimate articulatory features directly from speech signals~\cite{ toda2004acoustic}.

In recent years, deep learning algorithms have become popular state-of-the-art methods for AAI~\cite{siriwardena2022acoustic, wang2022acoustic, sun2021temporal, uria2012deep, shahrebabaki2021raw, bozorg2021autoregressive, maharana2021acoustic, udupa2021estimating}. Since these methods are data-driven and depend on the limited amount of articulatory data, they are unable to sufficiently generalize to unseen speakers to our knowledge. In this work, we aim to help bridge this gap through improving the AAI deep learning framework. Specifically, we examine different architecture configurations, features and training objectives to showcase a suitable framework that improves the state-of-the-art by 12.5\% using autoregression, adversarial training, and self supervision. We also evaluate predicted features through the articulatory phonology lens, comparing phoneme-level articulatory trajectories with expected human speech production behavior. Analyzing trajectories this way allows us to perform evaluation on any speaker, not just those who have recorded articulatory labels. Our analysis of estimated articulator movements also highlights the interpretability of our speech representations. Finally, we also evaluate inversion on speakers without articulatory labels using resynthesis and show that our inversion approach is also effective here. Code and additional related information are available at \href{https://github.com/articulatory/articulatory}{https://github.com/articulatory/articulatory}.

\section{Datasets}

\begin{figure}[t]
  \includegraphics[width=86mm]{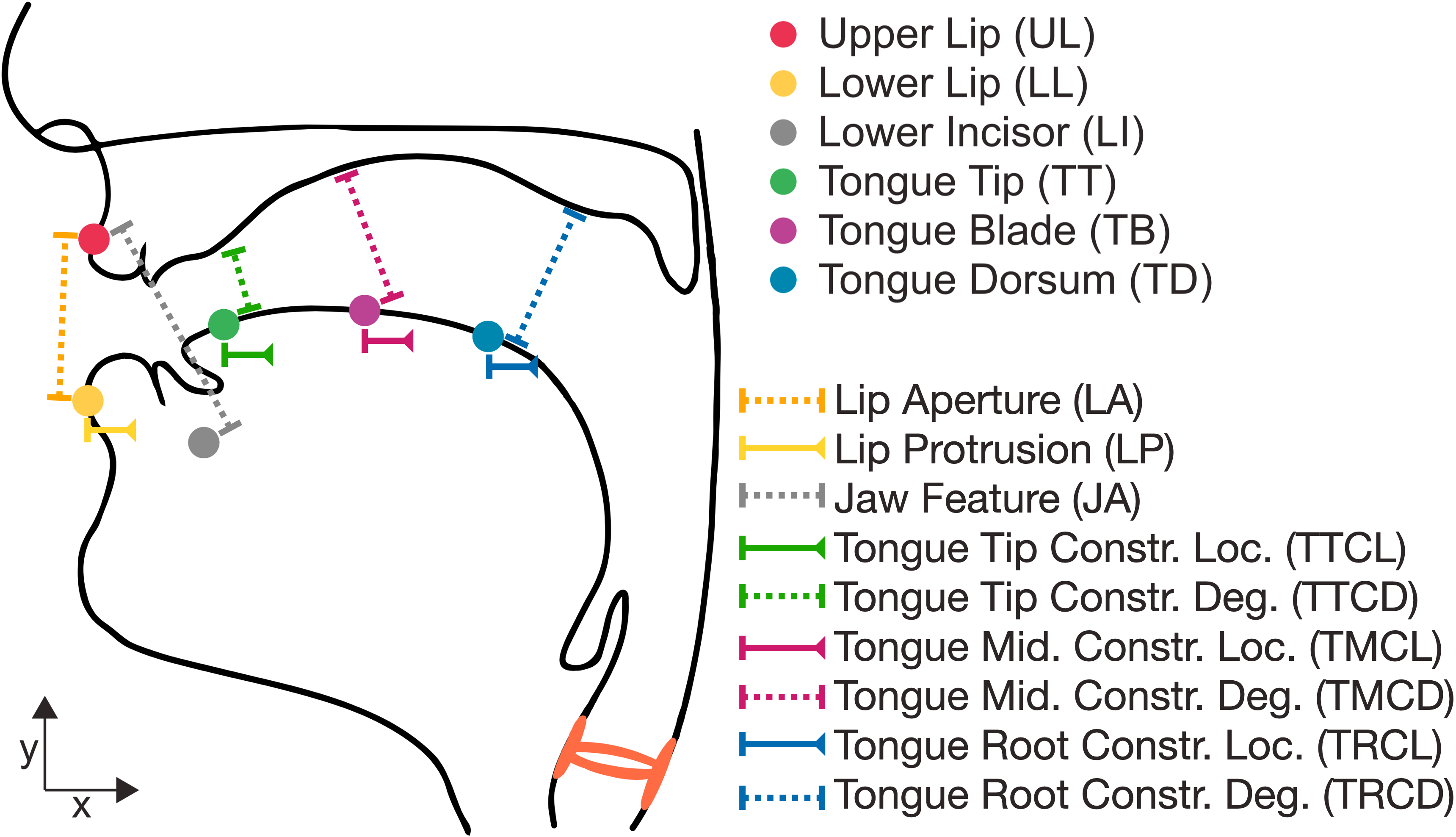}
  \caption{EMA features (points) and tract variables (segments). Diagram extends \cite{chartier2018encoding}. Details in Sections \ref{sec:mocha_dataset} and \ref{sec:tvs}.}
  \label{fig:midsagittal}
\end{figure}

\begin{figure}[t]
  \includegraphics[width=77mm]{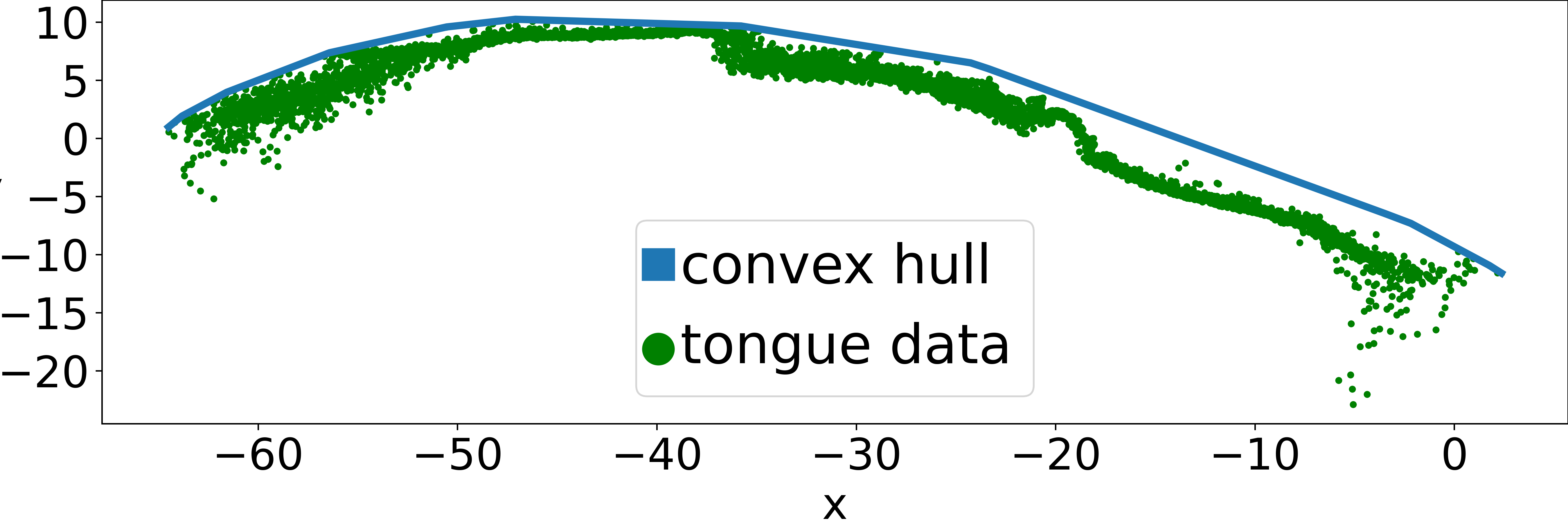}
  \caption{Estimating the palate location with a convex hull on tongue data for HPRC speaker F04. Details in Section \ref{sec:tvs}.}
  \label{fig:palate}
\end{figure}

\subsection{MOCHA-TIMIT EMA Dataset}
\label{sec:mocha_dataset}

We train our AAI models on an 8-speaker dataset using MOCHA-TIMIT~\cite{wrench1999mocha, wrench2000multi} and MNGU0~\cite{richmond2011mngu0}, containing 5.2 hours of 16 kHz speech. Data is collected with the speaker instrumented in an electromagnetic articulography (EMA) machine providing 200 Hz samples of EMA features. These 12-dimensional features contain the midsagittal x and y coordinates of jaw, lip, and tongue positions. Specifically, the 6 x-y positions correspond to the lower incisor, upper lip, lower lip, tongue tip, tongue body, and tongue dorsum, as visualized in Figure \ref{fig:midsagittal}. To study how well our AAI approach generalizes to unseen speakers, we arbitrarily choose MSAK as the unseen speaker, evaluating on all MSAK data and training on the other 7 speakers. Within the 7-speaker training set, we randomly set off 10\% of the utterances for the validation set.

\subsection{HPRC EMA Dataset}

We also follow previous works~\cite{seneviratne2019multi,9640504,9413742} to evaluate on the Haskins Production Rate Comparison (HPRC)~\cite{Tiede2017QuantifyingKA} database.
HPRC is an 8-speaker dataset containing 7.9 hours of 44.1 kHz speech and 100 Hz EMA. We downsampled the audio to 16kHz before further processing. As done in previous work~\cite{seneviratne2019multi}, we only consider information along the midsagittal plane and do not use the provided mouth left and jaw left data, matching our MOCHA-TIMIT EMA feature set described in Section \ref{sec:mocha_dataset}.

For our AAI train-val-test split, we hold out 1 male and 1 female speaker for our test set and train on the remaining 6 speakers, as done in~\cite{siriwardena2022acoustic}. While~\cite{siriwardena2022acoustic} also used these two speakers in their validation set, we put all of the data from both speakers in the test set in order to fully experiment within the unseen speaker setting. We note that our formulation increases the difficulty of the task compared to~\cite{siriwardena2022acoustic} since hyperparameter tuning would not lead to overfitting on the test speakers. However, our AAI method still obtains state-of-the-art Pearson correlation scores with this dataset, as detailed in Section \ref{sec:results}. For our train-val split, we use the same 90\%-10\% approach as the MOCHA-TIMIT split in Section \ref{sec:mocha_dataset} above.

\subsection{Combined EMA Dataset}

To train more generalizable models using a larger articulatory dataset, we combine MOCHA-TIMIT~\cite{wrench1999mocha, wrench2000multi, richmond2011mngu0} and HPRC~\cite{Tiede2017QuantifyingKA}, forming a 16-speaker EMA dataset. Here, we create a train-val-test split with 1000 val and 1000 test utterances, randomly assigning each utterance to one of the three sets. We note that all speakers appear in all three sets. Our resynthesis experiments use models trained on this combined dataset, discussed in Section \ref{sec:resynth}.

\subsection{Unseen Speaker Test Data without EMA Labels}
\label{sec:arctic}

We further evaluate our inversion approach using the CMU ARCTIC database~\cite{kominek2004cmu, wilkinson2016open}, an 18-speaker dataset composed of 14.3 hours of 16 kHz speech. In addition to the aforementioned held-out-speaker experiments, evaluations with this dataset also test our approach on unseen speakers. Moreover, ARCTIC contains various accents that are not present in our training set, testing the generalizability of our model to unseen accents. Since ARCTIC does not contain ground truth EMA labels, we evaluate our AAI model qualitatively by visualizing the predicted normalized lip aperture (LA) for individual vowels, detailed in Section \ref{ssec:vowel-art}. We conduct further AAI evaluations with this dataset quantitatively using a resynthesis approach discussed in Section \ref{sec:resynth}.

\section{Features}

\subsection{Tract Variables}
\label{sec:tvs}
Vocal tract positions in raw EMA are fairly speaker-dependent, which can be unsuitable for objective multi-speaker AAI evaluation~\cite{seneviratne2019multi}. Thus, we follow~\cite{seneviratne2019multi} to derive tract variables (TVs) that are more speaker-independent. As visualized in Figure \ref{fig:midsagittal}, we adopted 9 TVs: lip aperture (LA), lip protrusion (LP), tongue body constriction location (TBCL), tongue body constriction degree (TBCD), tongue tip constriction location (TTCL), and tongue tip constriction degree (TTCD).

Since one of our articulatory datasets does not contain tongue palate data, unlike~\cite{seneviratne2019multi}, we estimate the palate location. Specifically, we estimate the palate with a convex hull fitted on the tongue position data~\cite{wu2022art}. E.g., Figure \ref{fig:palate} plots the estimated palate locations for the MOCHA-TIMIT speaker MAPS. Then, we calculate the distance between the 3 EMA tongue positions (tip, body, dorsum) and the estimated palate as in~\cite{seneviratne2019multi}. All 9 TVs are calculated element-wise across time and thus have the same sampling rate as raw EMA features. Like~\cite{seneviratne2019multi}, we scale each speaker's TVs to between -1 and 1.

\subsection{Phonemic Features}
\label{sec:phonemic}

Since EMA data is limited, performing multi-task learning (MTL) with additional features can reduce overfitting~\cite{zhang2018mtl, zhang2018mtl}. 
Phonemic features implicitly contain articulatory information and properties not in EMA, e.g., voicing, and thus could improve AAI performance through MTL.~\cite{seneviratne2019multi} showed that adding a task of classifying aligned IPA-based phoneme labels provided in the dataset improves inversion quality. We extend this MTL approach to the fully unseen speaker setting with our baseline in this work.

Since AAI is typically formulated as a regression task, we also explore phonemic features that can be estimated in this manner. Specifically, we map each phoneme to an 18-dimensional vector, where each dimension corresponds to a different (place, manner) pair and is a binary value denoting the presence of the respective phoneme type, as in~\cite{anumanchipalli2019speech}. For all phonemic features, we use the same sampling rate as the EMA data in the respective datasets.

\subsection{Self-supervised Learning Features}
\label{sec:self_supervised}

Recent works~\cite{yang21c_interspeech} explored using pretrained self-supervised learning (SSL) speech features in speech processing tasks.
These features, trained on large-scale unsupervised speech corpora, have been shown to capture linguistic information~\cite{hsu2021hubert}. The use of SSL features in articulatory tasks is still relatively unexplored. To help bridge this, we use the final layer of HuBERT~\cite{hsu2021hubert, yang21c_interspeech}, a recent SSL model that has achieved state-of-the-art speech recognition performance.~\footnote{We used the hubert\_large\_ll60k model in \href{https://github.com/s3prl/s3prl}{https://github.com/s3prl/s3prl}.}

\subsection{Speaker Embeddings}
\label{sec:speaker_emb}

We note that the aforementioned features lack lots of speaker information, since the EMA-based features are normalized and the phoneme-based ones are effectively non-acoustic. Moreover, EMA- and phoneme-based features inherently lack speaker attributes like pitch~\cite{wu2022art}. Thus, to help our synthesis model generate speech more like the target speaker, we add an ECAPA-TDNN speaker embedding~\cite{speechbrain} to the input.

\section{Models}

\subsection{Articulatory Inversion Model}
\label{sec:inversion_model}

For our baseline AAI model, we use the bidirectional-GRU-based architecture in~\cite{siriwardena2022acoustic}. Specifically, this model is comprised of a two-layer bidirectional GRU~\cite{chung2014empirical} with size-256 hidden states followed by a two-layer multi-layer perceptron (MLP)~\cite{rumelhart1986mlp} with 128 hidden units. A probability-0.3 dropout layer succeeds each layer, and a batch norm layer precedes the MLP output layer. The model outputs 50 dimensions, 9 for TVs and 41 for classifying phonemes, with Tanh applied to the TV dimensions. We optimize our baseline using the mean absolute error (MAE) loss on the TVs added to $0.5$ times the cross-entropy loss on the phonemes, as in~\cite{siriwardena2022acoustic}. We use the same normalized MFCCs as ~\cite{siriwardena2022acoustic} as inputs to our baseline.

For our proposed inversion model, we build on the baseline by making it use chunked autoregression~\cite{morrison2022cargan, wu2022art} and adversarial training~\cite{kong2020hifigan, wu2022art}. Specifically, we encode outputs using an MLP and concatenate the encodings to subsequent input~\cite{morrison2022cargan, wu2022art}, which could help with modelling articulation dependencies across time, e.g., coarticulation \cite{browman1992articulatory}. With adversarial training, we use a convolutional neural network (CNN) discriminator to encourage the model to output more realistic estimates~\cite{kong2020hifigan, wu2022art}. We observed that using MAE with our 18-dimensional phonemic features, as introduced in Section \ref{sec:phonemic}, outperformed cross-entropy with one-hot phonemes here, potentially since the former let the discriminator more easily compare estimates with ground truths. Finally, we use the HuBERT features in Section \ref{sec:self_supervised} as inputs, linearly interpolated to match the TV sampling frequencies.

\subsection{Articulatory Synthesis Model}
\label{sec:synthesis_model}

For our resynthesis experiments in Section \ref{sec:resynth}, we use HiFi-CAR~\cite{wu2022art}, a generative model composed of residual CNN layers, to synthesize waveforms directly from our TV articulatory features. In order to provide the synthesizer with speaker information, we also concatenate the speaker embedding in Section \ref{sec:speaker_emb} to each time step in the input.

\begin{table}[t]
\centering
\begin{tabular}{ccc}
\hline
\textbf{Model} & \textbf{MOCHA} & \textbf{HPRC} \\
\hline
MFCC & $0.475$ & $0.677$ \\
MFCC, MTL (Baseline) \cite{siriwardena2022acoustic} & $0.502$ & $0.697$ \\
HuBERT & $0.641$ & $0.770$ \\
HuBERT, MTL & $\mathbf{0.678}$ & $0.782$ \\
HuBERT, MTL, AR, GAN & $0.672$ & $\mathbf{0.784}$ \\
\hline
\end{tabular}
\caption[Pearson correlation coefficients (PCCs) for held-out-speaker AAI. Models use multi-task learning (MTL), autoregression (AR) and adversarial training (GAN).]{
\label{tab:inversion} 
Pearson correlation coefficients (PCCs) for held-out-speaker AAI. Models use multi-task learning (MTL), autoregression (AR) and adversarial training (GAN).~\footnotemark
}
\end{table}

~\footnotetext{MOCHA results are updated from ICASSP 2023 version after correcting a preprocessing error.}

\begin{table*}[t]
\centering
\begin{tabular}{cccccccccccccccccc}
\hline
{} & \textbf{AA} & \textbf{AO} & \textbf{AW} & \textbf{ER} & \textbf{F} & \textbf{K} & \textbf{L} & \textbf{M} & \textbf{N} & \textbf{NG} & \textbf{OY} & \textbf{P} & \textbf{R} & \textbf{TH} & \textbf{V} \\
\hline
{\cite{siriwardena2022acoustic}} & $.206$ & $.214$ & $.229$ & $.211$ & $.208$ & $.222$ & $.222$ & $.216$ & $.195$ & $.244$ & $.226$ & $.206$ & $.206$ & $.207$ & $.198$ \\
{Ours} & $.175$ & $.182$ & $.187$ & $.168$ & $.163$ & $.193$ & $.177$ & $.176$ & $.163$ & $.213$ & $.191$ & $.178$ & $.173$ & $.179$ & $.165$ \\
{Diff.} & $.031$ & $.032$ & $\mathbf{.042}$ & $\mathbf{.042}$ & $\mathbf{.045}$ & $.029$ & $\mathbf{.045}$ & $\mathbf{.041}$ & $.032$ & $.031$ & $.035$ & $.028$ & $.033$ & $.028$ & $.033$ \\
\hline
\end{tabular}
\caption{\label{tab:l1}
Average L1 distances (\textcolor{ForestGreen}{$\downarrow$}) between estimated and ground truth TVs per phoneme. Phonemes listed here have the largest difference between the L1 of our method and that of the baseline \cite{siriwardena2022acoustic}. Details in \ref{sec:inversion}.
}
\end{table*}

\section{Results}
\label{sec:results}

\subsection{Articulatory Inversion}
\label{sec:inversion}

Table \ref{tab:inversion} summarizes our AAI results, measured using Pearson correlation as done previously \cite{siriwardena2022acoustic}. Our approaches in the last two rows improve the SOTA performance on the HPRC dataset by $12.5\%$ and outperform the SOTA model by $0.176$ correlation on MOCHA-TIMIT. 
Switching from MFCC inputs to HuBERT ones yields the largest improvement, suggesting that self-supervised features are better than spectral ones for multi-speaker AAI tasks. We attribute the lower performance on MOCHA-TIMIT compared to HPRC for all models due to HPRC having more hours of speech and potentially higher-quality EMA labels. This performance difference is probably not due to phoneme label quality, as the phoneme classification accuracy on MOCHA, 0.682, is comparable to that on HPRC, 0.673.

To further study which types of inputs our model contributes improved inversions, we compare the baseline with our proposed method at the phoneme level. Specifically, we calculate the average L1 distance between the predicted and ground truth tract variables, defined in Section \ref{sec:tvs}, in our HPRC test set for each phoneme. We obtain phoneme alignments with the Montreal Forced Aligner \cite{McAuliffe2017MontrealFA} and use the L1 distance metric since it can be computed at the frame level, e.g., as opposed to a correlation-based metric. Table \ref{tab:l1} summarizes these results. We observe that our model outperforms the baseline noticeably on nasals (M, N, NG) and liquids (L, R). One potential reason for this is that modelling improvements may help compensate for the lack of voicing and velar information in EMA. Generally, we observed that our model performs better than the baseline across all phonemes, and we encourage readers to see the supplementary material for all phoneme results.

\begin{figure}[t]
  \includegraphics[width=77mm]{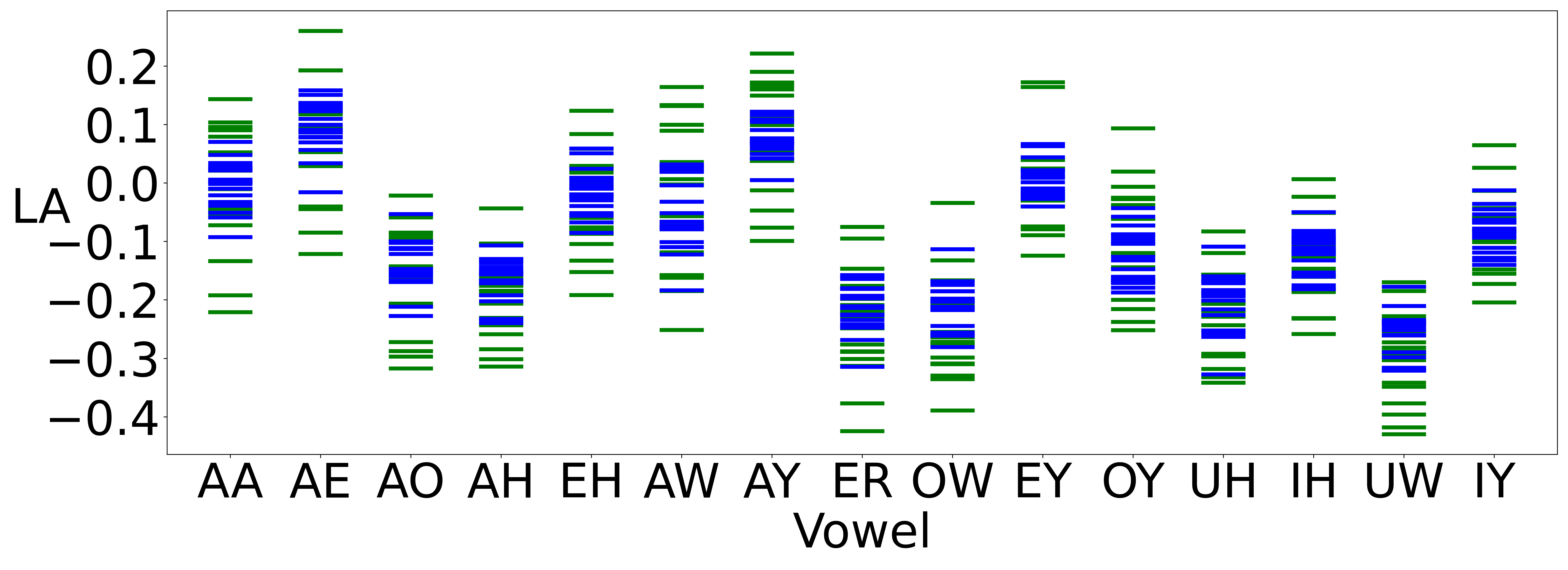}
  \caption{Predicted normalized lip aperture (LA) for our approach (blue) and the baseline (green) across 18 ARCTIC speakers for each vowel.}
  \label{fig:la}
\end{figure}

\subsection{Inversion Analysis without EMA Labels}
\label{ssec:vowel-art}

Figure \ref{fig:la} contains the predicted normalized lip aperture (LA) for our approach and the baseline across 18 ARCTIC speakers for each vowel. As mentioned in Section \ref{sec:arctic}, all of these speakers are unseen during training. Both approaches are able to generally predict the biologically plausible LA values. For example, the lip aperture, visualized in Figure \ref{fig:midsagittal}, is predicted to be wide for the "AE" sound and narrow for the "UW" sound. However, our model is much more consistent across speakers, as evinced by the lower variance. We note that this lower variance is consistent with the fact that the TVs are normalized and much less speaker dependent than EMA features~\cite{seneviratne2019multi}. Since the baseline incorrectly estimates lip aperture for a larger number of speakers, this suggests that our method is better at generalizing to unseen speakers.

\subsection{Resynthesis Analysis without EMA Labels}
\label{sec:resynth}

We can also compare inversion performance through resynthesizing speech from estimated features (Section \ref{sec:synthesis_model}) and evaluating synthesis quality. To study performance on unseen speakers this way, we resynthesize ARCTIC utterances using our synthesis model and each inversion method. Then, we compute the DTW-MCD~\cite{watanabe2018espnet} between the predicted and ground truth waveforms. Table \ref{tab:mcd} summarizes these results. Our approach outperforms the baseline for all speakers, consistent with our inversion results in Table \ref{tab:inversion} and earlier single-speaker experiments \cite{richmond2013evaluation}. We note that the MCD values are high since we did not optimize for synthesis quality in this work. Given the nascent nature of the multi-speaker articulatory synthesis direction, we plan to improve such models and continue validating inversion performance in this manner using MCD and more synthesis metrics~\cite{watanabe2018espnet,hayashi2021espnet2} moving forward.

\begin{table}[t]
\centering
\begin{tabular}{cccc}
\hline
\textbf{Model} & \textbf{AWB} & \textbf{SLT} & \textbf{Average} \\
\hline
Baseline \cite{siriwardena2022acoustic} & $9.23$ & $10.24$ & $10.94 \pm 0.76$ \\
Ours & $\mathbf{8.03}$ & $\mathbf{9.69}$ & $\mathbf{9.52 \pm 0.59}$ \\
\hline
\end{tabular}
\caption{\label{tab:mcd}
MCDs (\textcolor{ForestGreen}{$\downarrow$}) in resynthesis analysis experiments on ARCTIC speakers AWB and SLT, as well as the average and standard deviation across all speakers.
}
\end{table}

\section{Conclusion}

In this work, we devise an acoustic-to-articulatory inversion (AAI) approach in the context of generalizing to unseen speakers, improving state-of-the-art by 12.5\% on the HPRC task~\cite{Tiede2017QuantifyingKA}. We also show the interpretability of the estimated representations through directly comparing them with speech production behavior, evincing how this analysis can be done without labeled articulatory data. Finally, we propose a resynthesis-based AAI evaluation metric that does not rely on articulatory labels. We demonstrate the efficacy of this evaluation method using ARCTIC and observe results consistent with our inversion correlation metrics. In the future, we plan to extend our methodology to high-fidelity resynthesis and AAI for articulatory features beyond EMA.

\section{Acknowledgements}

This research is supported by the following grants to PI Anumanchipalli --- NSF award 2106928, Google Research Scholar Award, Rose Hills Foundation and Noyce Foundation.

\vfill\pagebreak

\bibliographystyle{IEEEbib}
{\small\bibliography{IEEEabrv,refs}}

\begin{thebibliography}{10}

\bibitem{lu2022characteristics}
Y.~Lu, C.~E. Wiltshire, K.~E. Watkins, et~al.,
\newblock ``Characteristics of articulatory gestures in stuttered speech: A
  case study using real-time magnetic resonance imaging,''
\newblock {\em Journal of Communication Disorders}, vol. 97, pp. 106213, 2022.

\bibitem{choi2021neural}
H.-S. Choi, J.~Lee, W.~Kim, et~al.,
\newblock ``Neural analysis and synthesis: Reconstructing speech from
  self-supervised representations,''
\newblock {\em NeurIPS}, 2021.

\bibitem{polyak21facebookresynthesis}
A.~Polyak et~al.,
\newblock ``{Speech Resynthesis from Discrete Disentangled Self-Supervised
  Representations},''
\newblock in {\em Interspeech}, 2021.

\bibitem{anumanchipalli2019speech}
G.~K. Anumanchipalli, J.~Chartier, and E.~F. Chang,
\newblock ``Speech synthesis from neural decoding of spoken sentences,''
\newblock {\em Nature}, vol. 568, no. 7753, pp. 493--498, 2019.

\bibitem{fant1991articulatorysynthesis}
G.~Fant,
\newblock ``What can basic research contribute to speech synthesis?,''
\newblock {\em Journal of Phonetics}, vol. 19, no. 1, pp. 75--90, 1991.

\bibitem{rubin1981articulatorysynthesis}
P.~Rubin, T.~Baer, and P.~Mermelstein,
\newblock ``An articulatory synthesizer for perceptual research,''
\newblock {\em The Journal of the Acoustical Society of America}, vol. 70, no.
  2, pp. 321--328, 1981.

\bibitem{scully1990articulatorysynthesis}
C.~Scully,
\newblock ``Articulatory synthesis,''
\newblock in {\em Speech production and speech modelling}, pp. 151--186.
  Springer, 1990.

\bibitem{wrench1999mocha}
A.~Wrench,
\newblock ``The mocha-timit articulatory database,'' 1999.

\bibitem{wrench2000multi}
A.~A. Wrench,
\newblock ``A multi-channel/multi-speaker articulatory database for continuous
  speech recognition research.,''
\newblock {\em Phonus.}, 2000.

\bibitem{richmond2011mngu0}
K.~Richmond, P.~Hoole, and S.~King,
\newblock ``Announcing the electromagnetic articulography (day 1) subset of the
  mngu0 articulatory corpus,''
\newblock in {\em Interspeech}, 08 2011, pp. 1505--1508.

\bibitem{Tiede2017QuantifyingKA}
M.~K. Tiede et~al.,
\newblock ``Quantifying kinematic aspects of reduction in a contrasting rate
  production task,''
\newblock {\em Journal of the Acoustical Society of America}, 2017.

\bibitem{zen2019libritts}
H.~Zen et~al.,
\newblock ``{LibriTTS}: A corpus derived from librispeech for text-to-speech,''
\newblock in {\em Interspeech}, 2019.

\bibitem{seneviratne2019multi}
N.~Seneviratne et~al.,
\newblock ``Multi-corpus acoustic-to-articulatory speech inversion.,''
\newblock in {\em Interspeech}, 2019.

\bibitem{toda2004acoustic}
T.~Toda, A.~Black, and K.~Tokuda,
\newblock ``Acoustic-to-articulatory inversion mapping with gaussian mixture
  model,''
\newblock in {\em Eighth International Conference on Spoken Language
  Processing}, 2004.

\bibitem{siriwardena2022acoustic}
Y.~M. Siriwardena et~al.,
\newblock ``Acoustic-to-articulatory speech inversion with multi-task
  learning,''
\newblock {\em Interspeech}, 2022.

\bibitem{wang2022acoustic}
J.~Wang et~al.,
\newblock ``Acoustic-to-articulatory inversion based on speech decomposition
  and auxiliary feature,''
\newblock in {\em ICASSP}, 2022.

\bibitem{sun2021temporal}
G.~Sun, Z.~Huang, L.~Wang, and P.~Zhang,
\newblock ``Temporal convolution network based joint optimization of
  acoustic-to-articulatory inversion,''
\newblock {\em Applied Sciences}, 2021.

\bibitem{uria2012deep}
B.~Uria, I.~Murray, S.~Renals, and K.~Richmond,
\newblock ``Deep architectures for articulatory inversion,''
\newblock in {\em Interspeech}, 2012.

\bibitem{shahrebabaki2021raw}
A.~S. Shahrebabaki, S.~M. Siniscalchi, and T.~Svendsen,
\newblock ``Raw speech-to-articulatory inversion by temporal filtering and
  decimation,''
\newblock {\em Interspeech}, 2021.

\bibitem{bozorg2021autoregressive}
N.~Bozorg, M.~T. Johnson, and M.~Soleymanpour,
\newblock ``Autoregressive articulatory wavenet flow for speaker-independent
  acoustic-to-articulatory inversion,''
\newblock in {\em SpeD}. IEEE, 2021.

\bibitem{maharana2021acoustic}
S.~K. Maharana, A.~Illa, R.~Mannem, et~al.,
\newblock ``Acoustic-to-articulatory inversion for dysarthric speech by using
  cross-corpus acoustic-articulatory data,''
\newblock in {\em ICASSP}. IEEE, 2021.

\bibitem{udupa2021estimating}
S.~Udupa, A.~Roy, A.~Singh, et~al.,
\newblock ``Estimating articulatory movements in speech production with
  transformer networks,''
\newblock {\em arXiv preprint arXiv:2104.05017}, 2021.

\bibitem{chartier2018encoding}
J.~Chartier, G.~K. Anumanchipalli, K.~Johnson, and E.~F. Chang,
\newblock ``Encoding of articulatory kinematic trajectories in human speech
  sensorimotor cortex,''
\newblock {\em Neuron}, 2018.

\bibitem{9640504}
A.~S. Shahrebabaki, G.~Salvi, T.~Svendsen, and S.~M. Siniscalchi,
\newblock ``Acoustic-to-articulatory mapping with joint optimization of deep
  speech enhancement and articulatory inversion models,''
\newblock {\em TASLP}, vol. 30, pp. 135--147, 2022.

\bibitem{9413742}
A.~S. Shahrebabaki, N.~Olfati, A.~S. Imran, et~al.,
\newblock ``A two-stage deep modeling approach to articulatory inversion,''
\newblock in {\em ICASSP}, 2021, pp. 6453--6457.

\bibitem{kominek2004cmu}
J.~Kominek and A.~W. Black,
\newblock ``{The CMU Arctic speech databases},''
\newblock in {\em Fifth ISCA workshop on speech synthesis}, 2004.

\bibitem{wilkinson2016open}
A.~Wilkinson, A.~Parlikar, S.~Sitaram, et~al.,
\newblock ``Open-source consumer-grade indic text to speech.,''
\newblock in {\em SSW}, 2016.

\bibitem{wu2022art}
P.~Wu, S.~Watanabe, L.~Goldstein, et~al.,
\newblock ``Deep speech synthesis from articulatory representations,''
\newblock in {\em Interspeech}, 2022.

\bibitem{zhang2018mtl}
Y.~Zhang and Q.~Yang,
\newblock ``An overview of multi-task learning,''
\newblock {\em National Science Review}, 2018.

\bibitem{yang21c_interspeech}
S.~wen Yang, P.-H. Chi, Y.-S. Chuang, et~al.,
\newblock ``{SUPERB: Speech Processing Universal PERformance Benchmark},''
\newblock in {\em Interspeech}, 2021, pp. 1194--1198.

\bibitem{hsu2021hubert}
W.-N. Hsu, B.~Bolte, Y.-H.~H. Tsai, et~al.,
\newblock ``Hubert: Self-supervised speech representation learning by masked
  prediction of hidden units,''
\newblock {\em TASLP}, 2021.

\bibitem{speechbrain}
M.~Ravanelli, T.~Parcollet, P.~Plantinga, et~al.,
\newblock ``{SpeechBrain}: A general-purpose speech toolkit,'' 2021,
\newblock arXiv:2106.04624.

\bibitem{chung2014empirical}
J.~Chung, C.~Gulcehre, K.~Cho, and Y.~Bengio,
\newblock ``Empirical evaluation of gated recurrent neural networks on sequence
  modeling,''
\newblock {\em NeurIPS}, 2014.

\bibitem{rumelhart1986mlp}
D.~E. Rumelhart, G.~E. Hinton, and R.~J. Williams,
\newblock {\em Learning Internal Representations by Error Propagation},
\newblock MIT Press, 1986.

\bibitem{morrison2022cargan}
M.~Morrison, R.~Kumar, K.~Kumar, et~al.,
\newblock ``Chunked autoregressive gan for conditional waveform synthesis,''
\newblock in {\em ICLR}, April 2022.

\bibitem{kong2020hifigan}
J.~Kong, J.~Kim, and J.~Bae,
\newblock ``{HiFi-GAN}: Generative adversarial networks for efficient and high
  fidelity speech synthesis,''
\newblock in {\em NeurIPS}, 2020.

\bibitem{browman1992articulatory}
C.~P. Browman and L.~Goldstein,
\newblock ``Articulatory phonology: An overview,''
\newblock {\em Phonetica}, 1992.

\bibitem{McAuliffe2017MontrealFA}
M.~McAuliffe et~al.,
\newblock ``Montreal forced aligner: Trainable text-speech alignment using
  kaldi,''
\newblock in {\em Interspeech}, 2017.

\bibitem{watanabe2018espnet}
S.~Watanabe, T.~Hori, S.~Karita, et~al.,
\newblock ``{ESPnet}: End-to-end speech processing toolkit,''
\newblock in {\em Interspeech}, 2018.

\bibitem{richmond2013evaluation}
K.~Richmond, Z.~Ling, J.~Yamagishi, et~al.,
\newblock ``On the evaluation of inversion mapping performance in the acoustic
  domain,''
\newblock in {\em Interspeech}, 2013.

\bibitem{hayashi2021espnet2}
T.~Hayashi et~al.,
\newblock ``Espnet2-tts: Extending the edge of tts research,''
\newblock {\em arXiv preprint arXiv:2110.07840}, 2021.

\end{thebibliography}

\end{document}